# THE GENESIS OF HIPPARCHUS' CELESTIAL GLOBE

**Susanne M. Hoffmann**

*Independent Scholar, formerly: Excellence Cluster TOPOI*
*Hannoversche Str. 6, 10115 Berlin, Germany*
*(akademeia@exopla.net)*

**ABSTRACT**

The globe of Hipparchus is not preserved. For that reason, it has been a source of much speculation and scientific inquiry during the last few centuries. This study presents a new analysis of the data given in the commentary on Aratus' poem by Hipparchus, in comparison with other contemporary Babylonian and Greek astronomical data, as well as their predecessors in the first millennium and their successors up to Ptolemy. The result of all these studies are the following: i) although the data of Ptolemy and Hipparchus are undoubtedly correlated, it is certainly also wrong to accuse Ptolemy having simply copied and transformed it without correct citation; ii) although Hipparchus presumably observed most of his star catalogue with his own instruments, we cannot neglect Babylonian influences. Hipparchus was educated in Greek astronomy but, in his time, there are traces of Babylonian influences since at least two centuries. Since we are unable to definitely prove that Hipparchus used Babylonian data, we are not sure if there are direct Babylonian influences in his time or as a consequence of his education only. Finally, we present a virtual 3D-image showing what the globe of Hipparchus might have looked like.

## 1. INTRODUCTION

This presentation is a summary of some of the results of the book on Hipparchus' celestial globe (Hoffmann, 2017) which appeared recently, but due to university laws had to be written in German. In this article, I would like to provide access to the book's information for the international community. The book contains four chapters: The first, of course, is a summary of the state of the art before my research, the second and third chapter describe the reconstruction of Hipparchus' star catalogue and analysis of this data with the aim of the virtual reconstruction of the globe. The fourth chapter presents numerous texts of different genres from both Babylonian and Greek astronomy including comparisons of their accuracy, style and data format with the equivalent in Hipparchus' text. The fifth chapter is a summary telling the (hi)story with the facts derived from the above research. In this article, I focus on the second chapter's reconstruction and the results of its analysis.

## 2. QUESTIONS AND AIMS

For nearly 400 years it has been known that Ptolemy's ecliptical longitudes have a systematic error. For at least 200 years, there are accusations that Ptolemy stole data from Hipparchus and transformed it with inaccurate precession. Although Ptolemy refers to Hipparchus' globe and cites data and constellation details from it, the Hipparchian globe cannot be completely reconstructed. There is a lack of information for some details and any reconstruction will only show a subset of the astronomical knowledge of Hipparchus.

However, the desire to build a model of the Hipparchian globe cannot be accomplished by simply using the Almagest star catalogue without any changes because Ptolemy in some issues reports deviations, e.g. "The star over the head, which Hipparchus [calls] 'the one on the muzzle'".[1] In some cases, he mentions deviations, but in other cases, he describes an asterism differently without mentioning any changes, e.g. the constellation of the crater has two handles and one star in the foot for Ptolemy, while for Hipparchus it does not have any handle and the foot contains four stars. Thus, Ptolemy's star catalogue is not a simple copy of Hipparchus' catalogue and we need an independent reconstruction of Hipparchus's catalogue to trace its roots and further development.

For the central question on the genesis of Hipparchus's globe we, therefore, have to pursue two general directions. First, how it is possible for us to reconstruct the historical (but missing) object, and second, how the historical object had been made. Did Hipparchus really observe everything by himself (which is generally assumed) or did he – like Ptolemy – also use Babylonian data in addition?

### 2.1. *Historical Questions*

The questions of the historian are related to the process of transfer and transformation of data, measuring and computing methods, as well as concepts (like frames of reference, metrical units, worldview, etc.) of uranometry in ancient science. It is known that Ptolemy used independent lists of longitudes (meridians) and lists of latitudes (*klimata*) to combine them to geographical coordinates as (longitude, latitude) pairs. Consequently, one of the historical questions is whether Hipparchus did something similar in astronomy combining earlier information. According to Pliny, Hipparchus was the first person to compile a star catalogue, but from what we know about the development of knowledge and scientific standards and conventions, they all have predecessors. What are the predecessors of Hipparchus? Did he use only earlier Greek data (which is attested to in the *Almagest*, Alm. VII, 1) or did he also use foreign sources of data – for instance, Babylonian sources?

There are numerous questions about Hipparchus' own observations: What kind of instrumentation did he use? Did he apply Babylonian methods to achieve a similar accuracy or did he use the instruments Ptolemy describes in the Almagest? Or does his scheme and method differ from both?

### 2.2. *Archaeoastronomical Aims*

The archaeological part of the research project is the virtual reconstruction of the missing globe. We know that the globe of Hipparchus existed because it is mentioned in the *Almagest*. Ptolemy cites the idea of an exact globe from Hipparchus but in some lines Ptolemy has a view which differs from that of Hipparchus, e.g. concerning the interpretation of constellation depictions (quoted above). We conclude from this that the globe of Hipparchus might have looked similar but not identical to Ptolemy's globe. Thus, we will use Ptolemy's exact descriptions as a more developed (later epoch) version of Hipparchus' object. Using only the mentioned differences in the *Almagest* in addition to Hipparchus's own descriptions and comparing the descriptions of Ptolemy with the appropriate parts of Hipparchus's one leads to a proper reconstruction of the two globes and the history in between them.

---

[1] Toomer and Ptolemaios, 1984, p. 361.

## 3. METHOD

The method of this research is quickly summarised: First, we reconstruct the star catalogue of Hipparchus by evaluating the data from the preserved two sources. Second, we evaluate the resulting database with the common and suggested algorithms of computational astronomy.

The preserved sources do not compile orthogonal coordinates of stars: Instead, the data they include are given in relative positions and events at the horizon. The subset of alignment–stars in the *Almagest* (Alm. VII, 1) are constructed to form lines of three stars. Thus, the astrometrical information in this text is only the relative position of three or four stars to each other. Interestingly, some of those stars quoted from Hipparchus are not in Ptolemy's star catalogue. This does not only suggest that Ptolemy's catalogue is not a simple copy of Hipparchus's one but also suggests an external source for Hipparchus's alignments.

The one and only preserved document penned by Hipparchus, the commentary on Aratus' poem on the *Phaenomena*, gives a corrected and much more systematic, even though not poetic, version of risings and settings of constellations. Hence, the data in this text is in the best cases a right ascension or declination, but in most cases it is a very broad range at the eastern or western parts of the horizon. If, and only if, a star happens to be mentioned two or three times, it is possible to compute its coordinates from this given risings, settings and culminations. The algorithms had been previously applied e.g. by (Vogt, 1925) and (Graßhoff, 1990) but repeated and improved by us.

In the text, the smallest unit Hipparchus distinguishes is a half degree in longitude. The accuracy is, therefore, a quarter of a degree and in declination only a half degree. However, Hippachus's text does not list all stars in his catalogue or a random subset of them but a carefully selected subset of rising, setting or culminating stars with well–defined constrains. That is why the average deviation of the selected stars from their real position (measured by modern measurements) may differ from this (Figure 1).

Although any reconstruction will be a subset and only a part of Hipparchus' full star catalogue, we discussed the error bars in different parts of the texts and compared the deviations of Hipparchus' data from modern data with the deviation of Ptolemy's and Babylonian data. The results are visualised in 2D diagrams and 3D plots as well as a 3D virtual reconstruction of Hipparchus's globe.

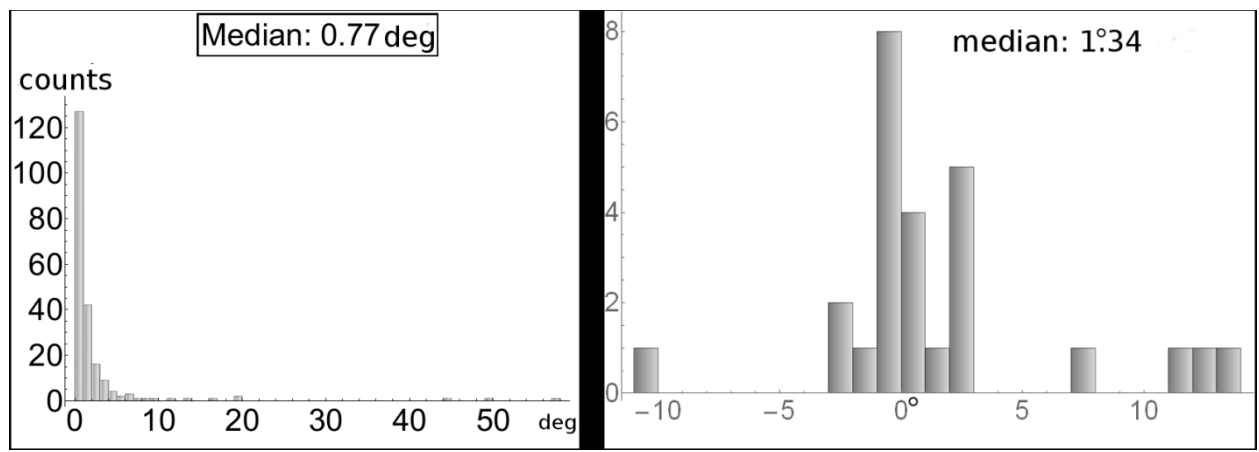

*Figure 1. Deviations of Hipparchus' right ascension from those in the HR catalogue: The data (left) in Hipparchus' second part with his own version of risings and settings has a smaller median error than the data (right) in the first part, where he discusses Aratus' verses. Does this indicate different sources?*

## 4. RESULTS

Testing the deviations of the reconstructed Hipparchian coordinates from the modern values we achieve different results in the direction of longitude and latitude and also different results for the different parts of his text. In the first part, where he lists his critique on Aratus' data Hipparchus tells us some declinations. Those declinations have a median deviation from the modern measurements by only 0°.088, while the right ascensions given in this part deviate by 1°.34 (median).[2] This average deviation for right ascensions is astonishingly big – larger than in the median deviation in the second part, where he achieved an average deviation of 0°.77 considering only the culminations or 0°.92 for all stars given.[3]

---

[2] See Figure 1 and for further information see (Hoffmann, 2017), pp. 194-211, especially p. 201 and p. 203.

[3] See Figure 1 and (Hoffmann, 2017), p. 118 (culminations) and p. 104 for all stars in the second part.

This is not a very good result but not surprising because his scale of his instrument allows him to distinguish only half degrees. Hence, the smallest possible average deviation is 0°.25λ on the ecliptic which transforms into different parts of °RA on the equator. However, with much bigger uncertainties in the observational data, this theoretically possible accuracy was maybe not achievable.

Concerning the huge differences between the first and second part of Hipparchus' text, we can suspect different sources. Could it be that the more accurate data has been observed directly, while the latter is read from a globe? This appears plausible, because Hipparchus might have denoted and published his own observations in a very systematic way, while the discussion of a poem verse–by–verse he probably did not during the night but on the desk. On the other hand, the declinations he gives in the first part have a surprisingly tiny median deviation (roughly 0°.1); is this only because of the small subset and the criteria of selection?

In the third part, Hipparchus defines 46 hour–stars by their right ascensions. Those right ascensions are also very accurate having a median deviation of 0°.16 from the modern measurement.[4] This might either also indicate an alternative source or be a result based upon much more observations than the other stars due to their importance or is caused by Hipparchus' selection. Choosing the median to average the data gives the most likely estimate for the errors of measurement because the biggest errors in our reconstructed data are probably not errors by Hipparchus but from two thousand years of copying tradition.

These results concerning the errors and deviations of Hipparchian stellar positions derived from the only preserved text make the accusation of Ptolemy being a fraud very suspicious: The globe of Hipparchus obviously had a scale with one or a half degree as the smallest unit while Ptolemy's star catalogue distinguishes 5 arcminutes as smallest unit. Additionally, we guess that Hipparchus' catalogue had about 800 stars out of which we can reconstruct 62 to 67 (depending on identification) full coordinates of declination and right ascension plus roughly 230 right ascensions (without declination) from culminating stars.

The original coordinate system of Hipparchus had very likely been the equatorial system:[5] Since the declinations are much more precise than the right ascensions, he probably used a different instrument and method to determine them: Since right ascensions are easily measurable by clocks (via siderial time) and ancient clocks had accuracies up to 20 minutes[6] this measurement might have been less accurate than a (probably more difficult) measurement of declinations with an angular instrument.

Measuring declinations and right ascensions while writing declinations and (apparent) ecliptical longitudes, Hipparchus used a different frame of reference for his observations and for his writing, and both systems are not the frame of reference of the *Almagest* (longitude, latitude). The latter is an orthogonal coordinate system with the ecliptic as baseline and Ptolemy in *Alm*. V, 1 even describes the construction of an instrument (a sort of spherical astrolabe) to measure those coordinates directly. However, Hipparchus at an earlier state of the art observed declinations and right ascensions separately with different methods and instruments.

### 4.1. *Archaeoastronomical Result*

With the data of the reconstructed part of a star catalogue, we can easily model a globe. The model presented here follows the description in the Almagest. Ptolemy instructs to build a (wooden) sphere, paint it dark and set the star–dots according to the description (coordinates, brightness, and colour) in the star catalogue, but generally in light colours on the dark background.

Concerning the other elements described in the building instructions of the Almagest, we should carefully select which of them we are sure about for Hipparchus' globe. Hipparchus' globe must have had a horizon because he reads risings and settings for his commentary. It must have had a meridian because he reads culminations and this wooden ring serves great for a detailed scale of declinations. He must have been able to easily figure out a 24th part of the equator to read the hour stars so the globe probably had a pattern of 24 right ascension–hours. Writing the commentary Hipparchus was not aware of precession.[7] Therefore, his globe had probably only one axis to turn around and not two axes like Ptolemy's globe. Nonetheless, we assume that Hipparchus placed the stars as exactly as described in the Almagest and afterwards somehow summarized the stars of the same constellation with a decent drawing or dashed borderline like it has been the astronomical tradition of globe makers until Joh. E. Bode in the 19th century.

The result of our reconstruction is displayed in **Figure 2**. The drawn declination lines are not sure and the polygons summarizing the constellations are the convex hulls of the reconstructed coordinates of the stars which are surely preserved: Since this is only a subset of Hipparchus' catalogue, certainly the constellation polygons are too small but this is the only data we have at the moment.

---

[4] Hoffmann, 2017, p. 122-135, especially p. 125.
[5] Duke, 2002, and Hoffmann, 2017, p. 612-643.

[6] Steele, 2000.
[7] Hoffmann, 2017, record 2.3.6 on p. 181.

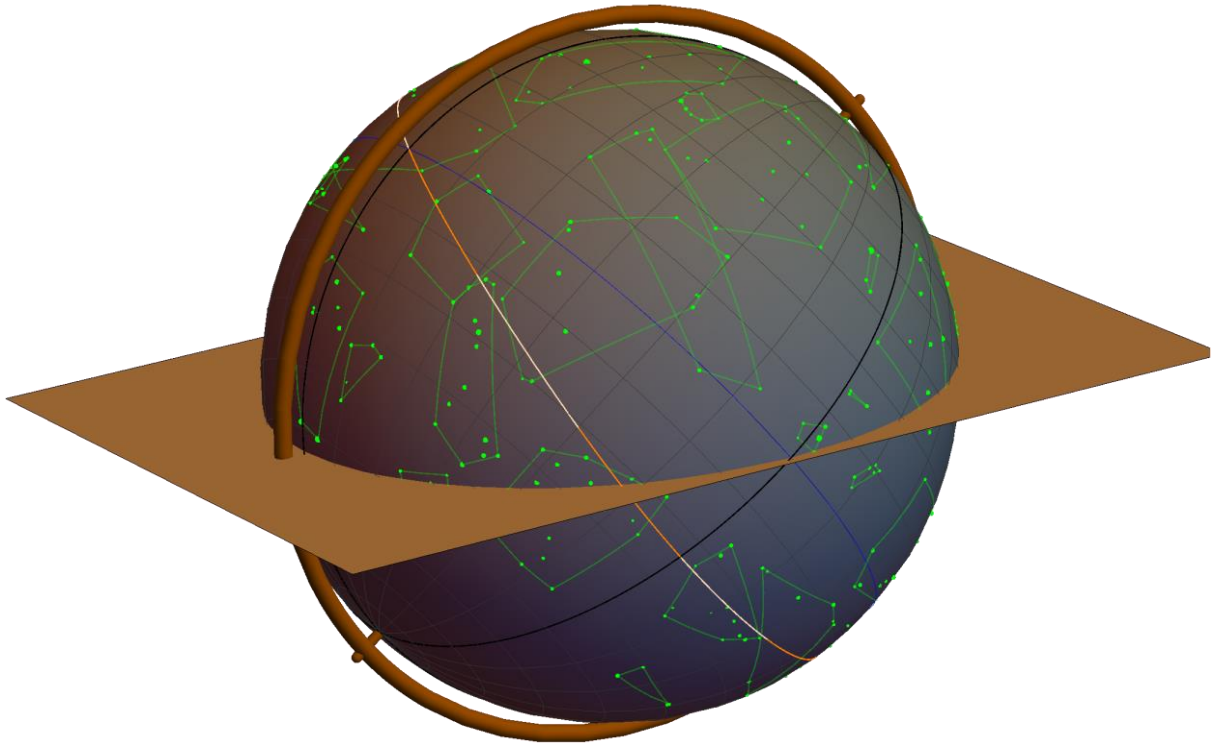

*Figure 2. This is how Hipparchus' globe roughly looked like: It had a horizontal plane and a meridian circle, the "longitude"–lines were 24 right ascension hours and the scale in declination has been either none or equally broad but there was a very detailed declination scale on the wooden meridian ring. The ecliptic – either drawn on the globe or modelled by another wooden ring – was divided into 12 signs times 30 parts, 360 in total. Whether or not the celestial equator was marked is not preserved.*

## 4.2. Historical Results

Aiming to answer the historical questions on the relation of Hipparchus's data to the Almagest and to Babylonian predecessors, we analysed the metrical units in which the data was given as well as the errors of the coordinates. Since, for example, the star pi Hydrae already had been a key witness (Graßhoff, 1990) for a correlation of the catalogues of Hipparchus and Ptolemy because it has the same error in both of them, we have been surprised by some important differences. The fact that the two catalogues differ and nevertheless are correlated suggests that Hipparchus's original works had been corrected and transformed before Ptolemy: It might have been one of the many sources but not the only one.

## 5. RELATION TO BABYLONIAN MATHEMATICAL ASTRONOMY

Babylonian astronomy in Hellenistic time considered the zodiac only and used an ecliptical coordinate system there. We do not know if the zones outside the zodiac played any role in that Late Babylonian astrometrical culture. There is no Babylonian star catalogue preserved. The list in MUL.APIN (which is far too old for our considerations here) which is often called "star catalogue" is not a catalogue but belongs to the text genre of commentaries. Hence, it is almost useless for any comparison with Hipparchus's style and contemporary Greek and Babylonian style.

Since Babylonians in Hellenistic epochs used the so–called normal stars[8] to give positions of planets as distances to those stars, the positions of normal stars must have been written down somewhere. Unfortunately, only thirteen lines of such a list are preserved in the fragments BM 36609+ and BM 46083: The list which is reconstructed by (Roughton, Steele, and Walker, 2004) using both tablets, gives only ecliptical longitudes and no latitudes of normal stars. The accuracy is 0°.25 and therefore higher than in Hipparchus' text but what we estimated as the possible maximum accuracy of Hipparchus' data source. Therefore, it was worth questioning possible correlations, but the correlation plots do not lead to any link between this Babylonian and the later Greek data.[9]

---

[8] "Normal" derives from the Latin word "norma" for "the measure".

[9] See Figure 3 and analysis in (Hoffmann, 2017), pp. 449-460, especially diagrams on p. 459.

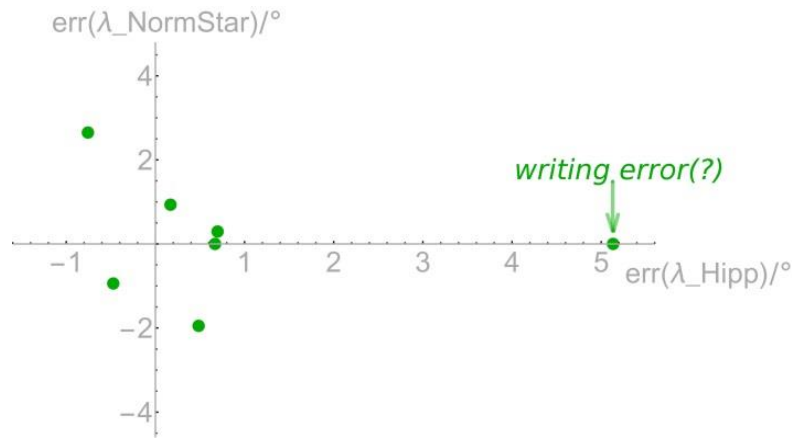

*Figure 3. Correlation plot for longitudes in degree of stars in Hipparchus' text and the Babylonian normal star lists.*

In Hipparchus' commentary, he does not use ecliptical coordinates. He thinks in equatorial coordinates giving declinations directly, and when he uses terms like "Leo 3°" this is a right ascension named after the degree of the ecliptic it passes.[10] That is why the hypothesis that Hipparchus might have directly used Babylonian data appears unlikely.

However, Hipparchus is not free from Babylonian influences. His circle is divided sexagesimally into 360 parts, he uses the Babylonian signs to divide the ecliptic and in the *Almagest* citations he uses the units "finger" and "cubit" for the tiny deviations from the lines which seems to be similar to the Babylonian units SI and KUŠ$_3$ respectively. The comparison of the real deviations and the values given by Hipparchus does not allow us to decide whether or not this hypothesis is true (50% chance). But in fact, this data leaves some room for interpretation.

However, the 2$^{nd}$ century BCE is known for some Greek innovations in mathematical astronomy. Maybe, Hipparchus learned from only Greek teachers who had been influenced by Babylonian astronomy directly or through the generations of teachers before. This hypothesis is very likely because Hipparchus cites an ecliptic divided in the same way as Eudoxus two centuries before him. That makes the Babylonian influence for Hipparchus only secondary and confirms the hypothesis of his own observations and developments in Greek mathematical astronomy. This result is especially interesting because Hipparchus' famous contemporary, Hypsikles, witnesses Babylonian influence on mathematical astronomy. Hence, we now found different development for computational and observational astronomy in the second century BCE.

## 6. RELATION TO PTOLEMY

Even the first plot of all Hipparchian stars on the same map as Ptolemy's stars shows that there are some stars mentioned by Hipparchus (at least in the alignments, but also in the commentary) that are missing in the *Almagest* star catalogue. So, if Ptolemy had stolen any data, why would he leave out some ten to twenty stars?

Considering the biggest errors in the reconstruction of Hipparchus' catalogue we find different positions than in Ptolemy's catalogue.[11] This is an allusion to a correction of Hipparchus' biggest errors before the writing of the *Almagest*. Considering that there are almost three hundred years of scientific progress between those two historical figures, we are allowed to assume some re-work of data in between. We suggest that "the ancient star catalogue" has not only been touched by Hipparchus himself and the school of Ptolemy writing the *Almagest*. The more likely model is a continuous usage and re-working of the data which we presume to have been originally observed by Hipparchus.

Ptolemy additionally used some Babylonian data, for instance, solar and lunar eclipse data. Since we concluded that Hipparchus likely did not have direct access to Babylonian data, this leads to the conclusion of new Babylonian influences in the time of Ptolemy or shortly before him.

A very likely model of the development is displayed in Figure 4.

The commentary had been written without considering precession. Later, Hipparchus discovered precession. Further generations of astronomers maybe did not observe the whole catalogue but checked and re-worked some apparent errors. It is still plausible that small stars without any practical meaning (like pi Hydrae) were never re-observed and, therefore, the error remained – even in the *Al-*

---

[10] I tested both possibilities but this one gives the better result (Hoffmann, 2017) and this is what e.g. (Duke, 2002) already found before me.

[11] Hoffmann, 2017, p. 618.

*magest* star catalogue – and, thus, the results of the dependency are still valid (Graßhoff, 1990). We only admit that most of the data is, of course, within the margin of error of ancient observations. Therefore, only roughly twenty stars of the ancient star catalogue analysed in (Graßhoff, 1990) are roughly laying on a line in the correlation plots (Graßhoff, 1990, pages 192–197) and, thus, allow the suggestion of a correlation. Out of the 1024 stars (plus five other objects) in the Almagest star catalogue, this is 1.7 to 1.8%. For the other stars no definite statement is possible. This small amount indicates a correlation and the rest is open for change due to a long tradition of working with the same catalogue rooting in Hipparchus' observational data and again supports our model of Figure 4.

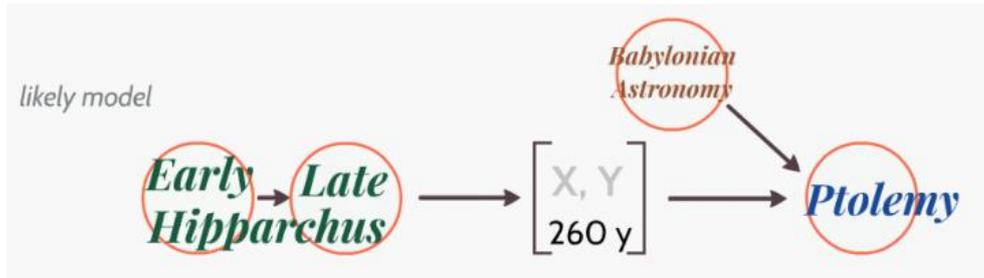

*Figure 4. A model for the most likely development of data from Hipparchus's commentary to Ptolemy's Almagest.*

## ACKNOWLEDGEMENTS

We thank the anonymous reviewers for their constructive comments. This work was a part of the Excellence Cluster TOPOI (Berlin), funded by the German Research Foundation (DFG). Special thanks to Gerd Graßhoff as a very helpful consultant.